\documentclass[english,aps,preprint,prb]{revtex4}
\usepackage[T1]{fontenc}
\usepackage[latin1]{inputenc}
\usepackage{float}
\usepackage{graphicx}

\makeatletter

\providecommand{\tabularnewline}{\\}

\usepackage{babel}
\makeatother
\begin{document}

\title{On the Dimers of Pseudoisocyanine }

\author{P.O.J. Scherer}

\email{philipp.scherer@ph.tum.de}

\affiliation{Physics Department T38, TU Munich, 85748 Garching}

\begin{abstract}
The self organisation of pseudoisocyanine-dimers in dilute aqueous
solutions is studied by classical MD simulations. The electronic structure
of the dimer is evaluated with the semiempirical ZINDO method to determine
the fluctuations of site energies and excitonic coupling. We study
different dimer conformations with blue or red shifted absorption
maxima as models for H and J-aggregates. The width of the absorption
bands is mainly explained by low frequency vibrations whereas the
fluctuations of site energies are less important. 
\end{abstract}
\maketitle

\section{introduction}

Since the discovery of the so called J-band \cite{Jelley,Scheibe},
an unusual sharp absorption band which is characteristic for the aggregation
of the classical sensitizing dye 1,1'-diethyl-2,2'-cyaninchloride
(pseudoisocyanine) a large amount of experimental and theoretical
work addressed the investigation of molecular aggregates and their
red shifted J-bands. At lower concentration blue shifted H-bands were
observed which were attributed to molecular dimers, the smallest possible
aggregates. Molecular modelling of the J-aggregates is difficult due
to the fact that the PIC molecule is a cation and therefore the Coulombic
interactions as well as dielectric shielding have to be taken into
account carefully. For the formation of larger aggregates the counterions
are important whereas this seems not the case for the smaller H-aggregates
\cite{Neumann}. Therefore we started the simulation of PIC aggregates
by a detailed investigation of the dimer. From the analysis of the
experimental spectrum in water \cite{Hallermeier} it was deduced
that both excitonic components contribute with an intensity ratio
of 2:1. This can not be explained\cite{Bird} by dimer models\cite{Hallermeier,Graves}
where the dipole moments are almost parallel or antiparallel as it
is the case for the common brickwork or ladder models which are found
in the literature for the J-aggregate\cite{Nolte,Scheibe2}. Another
focus of our investigations concerns the contribution of local Coulombic
interactions to the inhomogeneous broadening of the site energies
and its importance in comparison to intramolecular vibrations.

\section{methods}

For the classical MD simulation we used the model of rigid rotors
which can be easily combined with quantum calculations to obtain electronic
excitations and coupling matrix elements \cite{mycitation}. Since
also the position of the ethyl groups is fixed we have to distinguish
not only two stereoisomers but also a fully C2-symmetric form and
another form where the ethyl-groups break this symmetry. A possible
interconversion between these conformations was not taken into account.

We simulated a cube containing a pair of PIC (positively charged)
molecules and 2150 TIP5 \cite{TIP} water molecules. We did not use
periodic boundary conditions to avoid artefacts from the Coulombic
interaction with the mirror images. Instead reflecting boundaries
kept the molecules from escaping the box by reversing the normal velocity
component whenever the center of mass of one of the molecules encountered
the boundary . The boundary distance of 36\AA\quad was adjusted to
reproduce the experimental density of water at room temperature. The
equations of motion were solved using an implicit quaternion method
\cite{Quat} for the rotations and a Leap frog method for the translations.
The timestep used was 1 fsec. 

In our simulation we neglected electrostatic interactions with solvent
outside the cube. We calculated the missing contribution to the solvation
energy from a simple PCM model\cite{PCM}. The simulated box was put
into a cubic cavity in a dielectric continuum and the contribution
to the solvation energy was calculated from the interaction between
the charges within the box and the induced surface charges. It gave
13\% of the total solvation energy. This value stayed rather constant
along the trajectory. Therefore we assume that the essential changes
of interaction with solvent molecules in the immediate surroundings
are taken into account sufficiently.

The force field was designed to reproduce the local electrostatic
interactions properly, which is especially important for the large
sized PIC molecule with its extended $\pi$-electron system. It is
based on a simplified version of the effective fragment model \cite{EFP,H2O EFP,H2O EFP2}.
The charge distribution is approximated by distributed multipoles
which were calculated with GAMESS on the basis of TZV/HF wavefunctions\cite{Gamess}.
For the simulation only point charges $q_{i}$ and dipoles $\vec{p}_{i}$
were used which are centered at the positions of the nuclei and the
bond centers. The Coulombic interaction energy is

\begin{equation}
V_{ij}^{Coul}=\frac{q_{i}q_{j}}{4\pi\epsilon_{0}R_{ij}}+\frac{\vec{R}_{ij}(q_{i}\vec{p_{j}}-q_{j}\vec{p}_{i})}{4\pi\epsilon_{0}R_{ij}^{3}}+\frac{R_{ij}^{2}\vec{p_{i}}\vec{p}_{j}-3(\vec{R}_{ij}\vec{p}_{i})(\vec{R}_{ij}\vec{p}_{j})}{4\pi\epsilon_{0}R_{ij}^{5}}\end{equation}

The values of point charges and dipoles are given for the symmetry
unique atoms in Table$\,$\ref{table pic}. 

The electronic spectra were calculated on the ZINDO/CI-S level\cite{Zerner}
including the point monopoles and dipoles of all water molecules.
The two lowest excited states of the dimer are to a large extent linear
combinations of the lowest monomer excitations with only very small
admixture of higher monomer excitations and of charge resonance states.
Therefore we use a simple 2-state model to analyze the delocalized
dimer states in terms of the local excitations $A*\mbox{ and }B*$.
The interaction matrix is \begin{equation}
\left(\begin{array}{cc}
\overline{E}-\Delta/2 & V\\
V & \overline{E}+\Delta/2\end{array}\right)\end{equation}
with the average monomer transition energy $\overline{E}=(E_{A*}+E_{B*})/2$,
their splitting $\Delta=E_{B*}-E_{A*}$ and the excitonic coupling
$V$ . Its eigenvectors are the two delocalized dimer excitations
which are written with mixing coefficients $\cos\gamma,\sin\gamma$
as\begin{equation}
|1>=\cos\gamma|A*>+\sin\gamma|B*>\quad|2>=-\sin\gamma|A*>+\cos\gamma|B*>\end{equation}

The transition dipoles of these two states

\begin{equation}
\vec{\mu}_{1}=\cos\gamma\vec{\mu}_{A*}+\sin\gamma\vec{\mu}_{B*}\quad\vec{\mu}_{2}=\sin\gamma\vec{\mu}_{A*}-\cos\gamma\vec{\mu}_{B*}\label{tdip}\end{equation}

are linear combinations of the transition dipoles $\vec{\mu}_{A*,B*}$
of the two monomers which are assumed to be independent from the excitonic
interaction. Therefore the mixing angle $\gamma$ can be determined
from a least square fit of the two dimer transition dipoles to (\ref{tdip}).
Then the elements of the interaction matrix are calculated from the
transition energies of the two dimer excitations as

\begin{equation}
\overline{E}=\frac{E_{1}+E_{2}}{2}\end{equation}
 \begin{equation}
\Delta=(\cos(\gamma)^{2}-\sin(\gamma)^{2})(E_{2}-E_{1})\end{equation}
 \begin{equation}
V=\cos(\gamma)\sin(\gamma)(E_{2}-E_{1})\end{equation}

We want to emphasize that this analysis is based on the delocalized
dimer orbitals. It does not involve any kind of multipole expansion,
especially the calculated excitonic coupling is not of the dipole-dipole
type which would be quite questionable at such short intermolecular
distances. To check the quality of the ZINDO method, we compared the
results for a selected sandwich dimer configuration with a much more
elaborate 631G{*}{*} HF/CI calculation. The calculated transition
dipoles of the two lowest singlet excitations were very similar (3
and 16 Debyes from ZINDO , 2 and 14 Debyes from HF/CI ), the excitonic
splitting was somewhat larger for the HF/CI method (0.54eV as compared
to 0.37eV for ZINDO). Both methods placed the lowest charge resonance
states at about 0.65 eV above the upper excitonic band.

\section{Results}

We determined the vibronic coupling parameters for the PIC monomer
as described in our earlier paper \cite{book}. Application of ab
initio methods \cite{Gamess} improved the quality of the results
so that a direct comparison with the profile of the absorption spectrum
becomes feasible. The normal modes were calculated on the 6-31G/MP2
level and the coupling to the optical transition on the CI/SD level.
Using these couplings and the displaced harmonic oscillator model
the lineshape was calculated as the Fourier transformed time correlation
function. In the low frequency region the largest vibronic couplings
are found for normal modes at 40 and 46 $cm^{-1}$ which contribute
significantly to the broadening of the absorption band. Another important
contribution from modes around 1500 $cm^{-1}$ which are also known
from Raman spectra is the origin of the observed vibrational progression.
Further modes between 50 and 1400 $cm^{-1}$ form a rather dense continuum
of coupling states. The simulated spectrum (fig. \ref{Picabso}) largely
resembles the experimental absorption profile. The width of the simulated
bands is somewhat too small and the intensity of the prominent stretching
modes is slightly overestimated. This could be possibly further improved
by taking frequency changes and mode coupling into account. 

The MD simulations were started from several plausible dimer structures.
First the PIC molecules were kept fixed and the solvent was equilibrated
for 50 psec. Then the restraints were removed and the system was simulated
for another 50 psec. The distance and orientation of the two PIC molecules
were analyzed to identify periods of relative stability.

Starting from a sandwich structure, a rather stable structure evolved
within 10 psec (fig.3a). It is not symmetric but still there is almost
no splitting of the calculated site energies (Table 2) which show
rapid fluctuations with components down to 20fsec. Such fast fluctations
are well known from experimental and theoretical work on the dynamics
of dephasing and solvation in molecular liquids\cite{Nibbering}.
They have been attributed to the inertial motion of the solvent molecules,
which show up as the Gaussian shaped rapid initial decay of the solvation
time correlation function\cite{Maroncelli,Carter}. In our simulations
the orientational time correlation function of the water molecules
can be described by a Gaussian with a correlation time of 60 fs at
short times. The time correlation of the electrostatic potential decays
faster. The initial Gaussian decay with 20 fs is very similar to that
of the correlation function of the transition energies. Probably collective
librational motions contribute more efficiently to the electronic
dephasing than the motion of the individual molecules.\cite{Nibbering,Mukamel} 

The center of the site energies is shifted by 0.06 eV to lower energies
as compared to a monomer in vacuum and the variance of the site energies
is comparable to that of a monomer. The two transition dipoles are
almost parallel and the lower transition carries only 2\% of the total
oscillator strength. The excitonic coupling shows fluctuations similar
to the site energies. Its variance, however, amounts to only 7\% of
the average value of 0.25eV. Hallermeier et al \cite{Hallermeier}
deduced a smaller excitonic coupling of 0.078 eV. Most probably their
dimer spectrum has some admixture of the monomer spectrum. We assume
that the absorption maximum at 520nm is due to monomers and the maximum
of the real dimer spectrum is at 480nm. This would be consistent with
an H-aggregate with an excitonic coupling of 0.2eV.

We studied also brickwork structures as a model for the J-aggregates
with a red shifted absorption. We found a relative stable structure
which is shown in fig. 3b. The structural fluctuations are much larger
than for the sandwich model but the fluctuations of site energies
and excitonic coupling are even somewhat smaller. The coupling of
-0.064eV is close to the value of -0.078eV which was used to simulate
the vibronic spectrum of the J-aggregates\cite{book}. 

\begin{acknowledgments}
This work has been supported by the Deutsche Forschungsgemeinschaft
(SFB 533)
\end{acknowledgments}

\newpage

\section{Table Captions}

\subsection*{\noindent Table \ref{table pic}: }

\noindent coordinates, atomic charges and dipoles for PIC

\subsection*{\noindent Table \ref{tablestat}:}

\noindent mean values and standard deviation of distances, orientation
angles, excitation energies and excitonic couplings for the two structures.
The long axis is defined by the vector connecting the two nitrogen
atoms $\mathbf{R}(N_{13})-\mathbf{R}(N_{14})$, the short axis by
the vector $\mathbf{R}(N_{14})-\mathbf{R}(C_{20})+\mathbf{R}(N_{13})-\mathbf{R}(C_{19})$.

\newpage

\begin{table}[H]

\caption{\label{table pic}}

\begin{center}\begin{tabular}{|c|c|c|c|c|c|c|c|}
\hline 
label &
x(Bohr)&
y(Bohr) &
z(Bohr) &
q &
$p_{x}(e\bullet Bohr)$&
$p_{y}$($e\bullet Bohr)$&
$p_{z}$($e\bullet Bohr)$\tabularnewline
\hline
C 1&
-9.565&
1.113&
-3.478&
0.973&
0.000&
0.044&
0.092\tabularnewline
\hline
C 3&
-7.152&
0.889&
-2.403&
0.929&
0.038&
-0.017&
0.029\tabularnewline
\hline
C 5&
-6.682&
1.814&
0.029&
0.928&
0.046&
0.017&
0.005\tabularnewline
\hline
C 7&
-8.685&
2.931&
1.361&
0.857&
-0.014&
-0.025&
-0.090\tabularnewline
\hline
C 9&
-11.028&
3.132&
0.263&
1.055&
0.105&
-0.062&
-0.039\tabularnewline
\hline
C11&
-11.494&
2.232&
-2.174&
0.992&
0.101&
-0.009&
0.026\tabularnewline
\hline
N13&
-4.247&
1.613&
1.055&
0.472&
-0.038&
-0.096&
-0.101\tabularnewline
\hline
C15&
-2.340&
0.409&
-0.158&
0.997&
-0.061&
0.124&
0.024\tabularnewline
\hline
C17&
-2.850&
-0.583&
-2.637&
0.738&
-0.052&
0.121&
0.041\tabularnewline
\hline
C19&
-5.129&
-0.334&
-3.709&
1.046&
0.019&
0.001&
0.116\tabularnewline
\hline
C21&
-3.738&
2.886&
3.487&
0.859&
-0.053&
-0.089&
-0.108\tabularnewline
\hline
C23&
-4.351&
1.265&
5.791&
0.893&
0.020&
-0.005&
-0.027\tabularnewline
\hline
C25&
0.000&
0.000&
1.050&
0.421&
0.000&
0.000&
-0.310\tabularnewline
\hline
H27&
-1.393&
-1.638&
-3.559&
0.393&
0.072&
0.038&
0.027\tabularnewline
\hline
H29&
-5.484&
-1.128&
-5.546&
0.341&
0.003&
-0.020&
-0.040\tabularnewline
\hline
H31&
-8.451&
3.626&
3.244&
0.331&
-0.016&
0.024&
0.046\tabularnewline
\hline
H33&
-12.532&
3.997&
1.319&
0.347&
-0.033&
0.016&
0.015\tabularnewline
\hline
H35&
-13.340&
2.406&
-2.997&
0.351&
-0.036&
0.003&
-0.016\tabularnewline
\hline
H37&
-9.869&
0.384&
-5.351&
0.331&
-0.028&
-0.008&
-0.037\tabularnewline
\hline
H39&
-1.778&
3.440&
3.479&
0.319&
0.024&
0.010&
-0.006\tabularnewline
\hline
H41&
-4.812&
4.618&
3.507&
0.313&
-0.023&
0.024&
-0.001\tabularnewline
\hline
H43&
-6.323&
0.721&
5.833&
0.299&
-0.033&
-0.015&
0.012\tabularnewline
\hline
H45&
-3.944&
2.325&
7.497&
0.315&
0.002&
0.013&
0.035\tabularnewline
\hline
H47&
-3.236&
-0.453&
5.827&
0.271&
0.003&
-0.036&
0.020 \tabularnewline
\hline
\end{tabular}\end{center}
\end{table}

\begin{table}[H]
\begin{center}\begin{tabular}{|c|c|c|c|c|c|c|c|}
\hline 
label &
x(Bohr) &
y(Bohr)&
z(Bohr) &
q &
$p_{x}(e\bullet Bohr)$&
$p_{y}$(Bohr)&
$p_{z}(e\bullet Bohr)$\tabularnewline
\hline
BO31 &
-8.358&
1.001&
-2.940&
-0.591&
-0.156&
0.084&
0.045\tabularnewline
\hline
BO53 &
-6.917&
1.351&
-1.187&
-0.737&
-0.091&
0.047&
-0.017\tabularnewline
\hline
BO75 &
-7.683&
2.372&
0.695&
-0.642&
-0.075&
-0.008&
0.003\tabularnewline
\hline
BO97 &
-9.856&
3.032&
0.812&
-0.820&
-0.051&
0.026&
-0.020\tabularnewline
\hline
BO111 &
-10.529&
1.672&
-2.826&
-0.889&
-0.029&
0.021&
0.054\tabularnewline
\hline
BO119 &
-11.261&
2.682&
-0.956&
-0.693&
0.049&
-0.074&
-0.055\tabularnewline
\hline
BO135 &
-5.464&
1.713&
0.542&
-0.348&
-0.128&
0.109&
-0.133\tabularnewline
\hline
BO1513&
-3.293&
1.011&
0.449&
-0.450&
0.093&
-0.118&
-0.042\tabularnewline
\hline
BO1715&
-2.595&
-0.087&
-1.398&
-0.495&
0.080&
-0.050&
-0.035\tabularnewline
\hline
BO193 &
-6.140&
0.277&
-3.056&
-0.488&
-0.014&
-0.194&
0.034\tabularnewline
\hline
BO1917&
-3.989&
-0.459&
-3.173&
-0.905&
0.001&
-0.013&
0.010\tabularnewline
\hline
BO2113&
-3.993&
2.250&
2.271&
-0.125&
0.086&
0.192&
0.103\tabularnewline
\hline
BO2321&
-4.045&
2.076&
4.639&
-0.341&
-0.016&
-0.003&
0.074\tabularnewline
\hline
BO2515&
-1.170&
0.204&
0.446&
-0.609&
-0.082&
-0.320&
0.223\tabularnewline
\hline
BO2625&
0.000&
0.000&
2.056&
-0.447&
0.000&
0.000&
-0.209\tabularnewline
\hline
BO2717&
-2.121&
-1.111&
-3.098&
-0.506&
-0.137&
0.205&
0.099\tabularnewline
\hline
BO2919&
-5.306&
-0.731&
-4.627&
-0.525&
0.037&
0.121&
0.276\tabularnewline
\hline
BO317 &
-8.568&
3.279&
2.303&
-0.537&
-0.046&
-0.114&
-0.263\tabularnewline
\hline
BO339 &
-11.780&
3.565&
0.791&
-0.547&
0.237&
-0.131&
-0.159\tabularnewline
\hline
BO3511&
-12.417&
2.319&
-2.586&
-0.551&
0.285&
-0.028&
0.126\tabularnewline
\hline
BO371 &
-9.717&
0.748&
-4.415&
-0.538&
0.061&
0.109&
0.278\tabularnewline
\hline
BO3921&
-2.758&
3.163&
3.483&
-0.555&
-0.314&
-0.072&
0.041\tabularnewline
\hline
BO4121&
-4.275&
3.752&
3.497&
-0.529&
0.200&
-0.282&
0.023\tabularnewline
\hline
BO4323&
-5.337&
0.993&
5.812&
-0.543&
0.350&
0.089&
-0.009\tabularnewline
\hline
BO4523&
-4.148&
1.795&
6.644&
-0.510&
-0.074&
-0.199&
-0.303\tabularnewline
\hline
BO4723&
-3.794&
0.406&
5.809&
-0.518&
-0.218&
0.302&
-0.004 \tabularnewline
\hline
\end{tabular}\end{center}
\end{table}

\newpage

\begin{table}

\caption{\label{tablestat}}

\begin{tabular}{|c|c|c|}
\hline 
&
sandwich&
brickwork\tabularnewline
\hline
\hline 
excitation energy $E_{a}$&
$2.522\pm0.014\, eV$&
$2.565\pm0.010\, eV$\tabularnewline
\hline 
excitation energy $E_{b}$&
$2.523\pm0.014\, eV$&
$2.561\pm0.010\, eV$\tabularnewline
\hline 
excitonic coupling $V_{exc}$&
$0.25\pm0.017\, eV$&
$-0.064\pm0.008\, eV$\tabularnewline
\hline 
center-center distance $R_{ab}$&
$4.25\pm0.08$\AA&
$8.33\pm0.24$\AA\tabularnewline
\hline 
angle between long axes&
$6.0\pm1.7^{o}$&
$34.0\pm4.0^{o}$\tabularnewline
\hline 
angle between short axes&
$174.0\pm2.6^{o}$&
$122.0\pm3.9^{o}$\tabularnewline
\hline 
distance of charge centers $R_{cc}$&
$5.42$\AA&
$7.57$\AA\tabularnewline
\hline
\end{tabular}
\end{table}

\newpage

\section{Figure Captions:}

\subsection*{\noindent \protect\begin{flushleft}Figure 1:\protect\end{flushleft}}

\noindent The atom numbering for PIC is shown as it is used to tabulate
the parameters

\subsection*{\noindent \protect\begin{flushleft}Figure 2:\protect\end{flushleft}}

\noindent \begin{flushleft}The experimental absorption spectrum \cite{Hallermeier}
of monomeric PIC (dots) is compared with a calculated spectrum (full
line) from the displaced oscillator model. The calculated spectrum
was shifted to reproduce the absorption maximum at 19100$cm^{-1}$\end{flushleft}

\subsection*{\noindent \protect\begin{flushleft}Figure 3:\protect\end{flushleft}}

\noindent \begin{flushleft}Starting from a sandwich (a) or brickwork
(b) structure relative stable dimer configurations evolved. The figure
shows representative snapshots. \end{flushleft}

\begin{figure}[H]

\caption{\label{numbers}}

\begin{center}\includegraphics[%
  scale=0.5,
  angle=270]{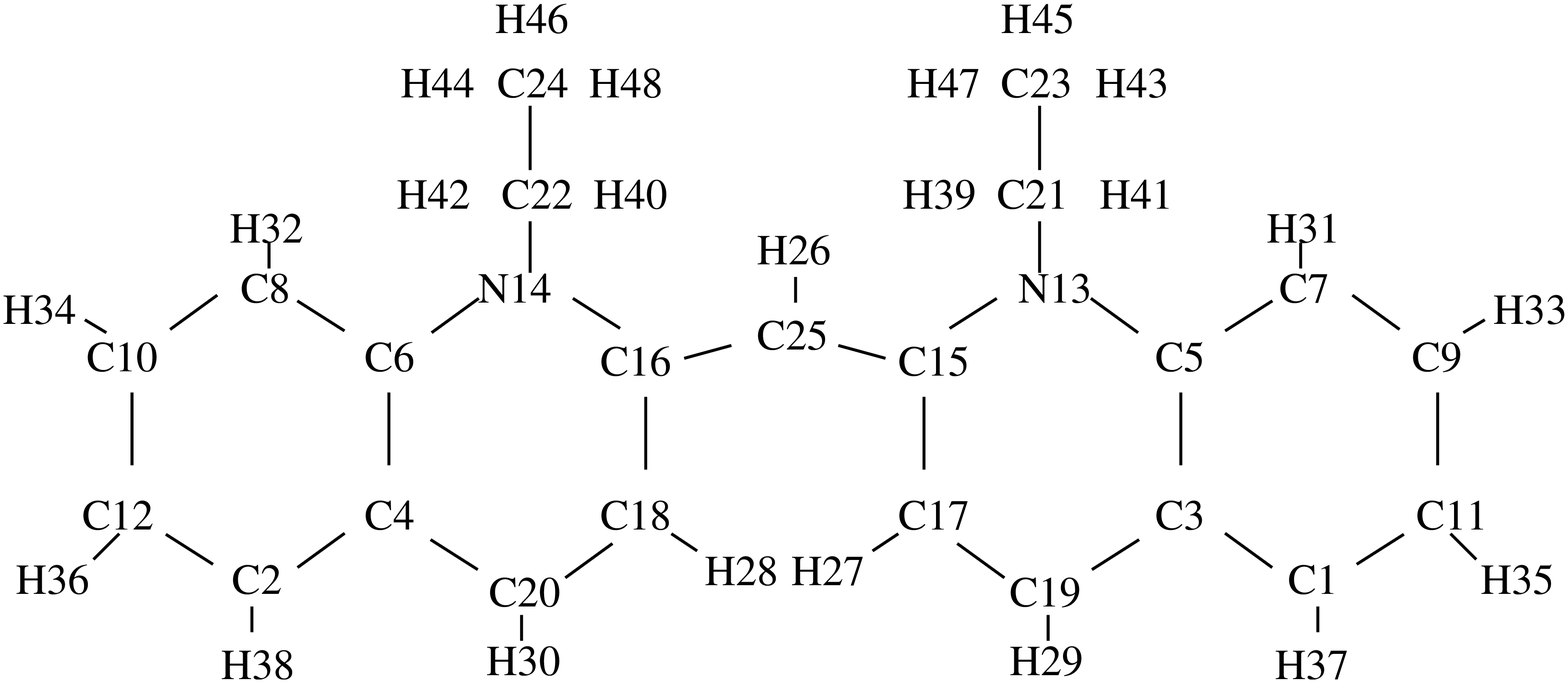}\end{center}
\end{figure}

\begin{figure}[H]

\caption{\label{Picabso}}

\begin{center}\includegraphics[%
  clip,
  scale=0.7,
  angle=270]{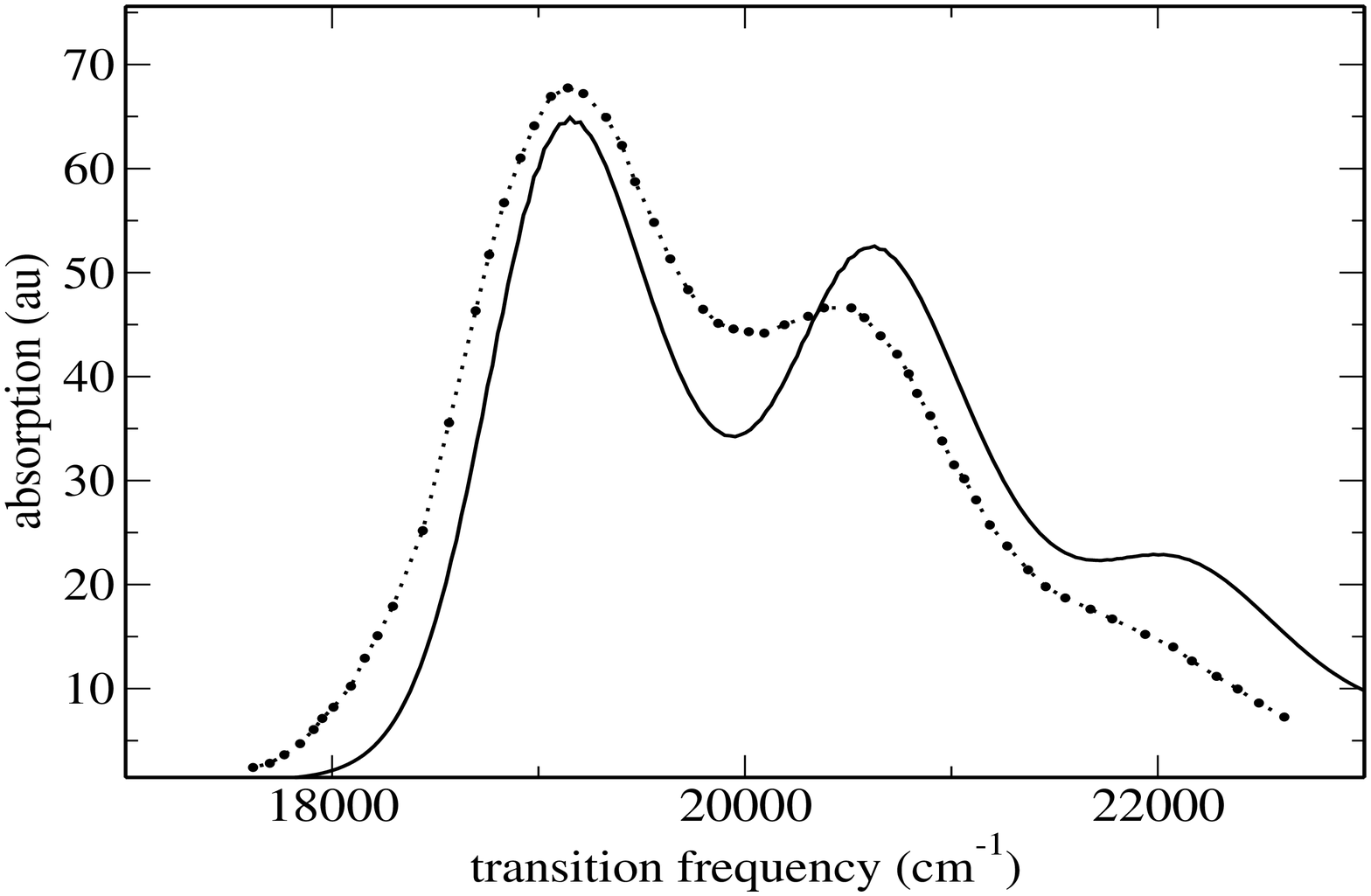}\end{center}
\end{figure}

\begin{figure}[H]

\caption{\label{structures}}

\begin{center}\includegraphics[%
  scale=0.6]{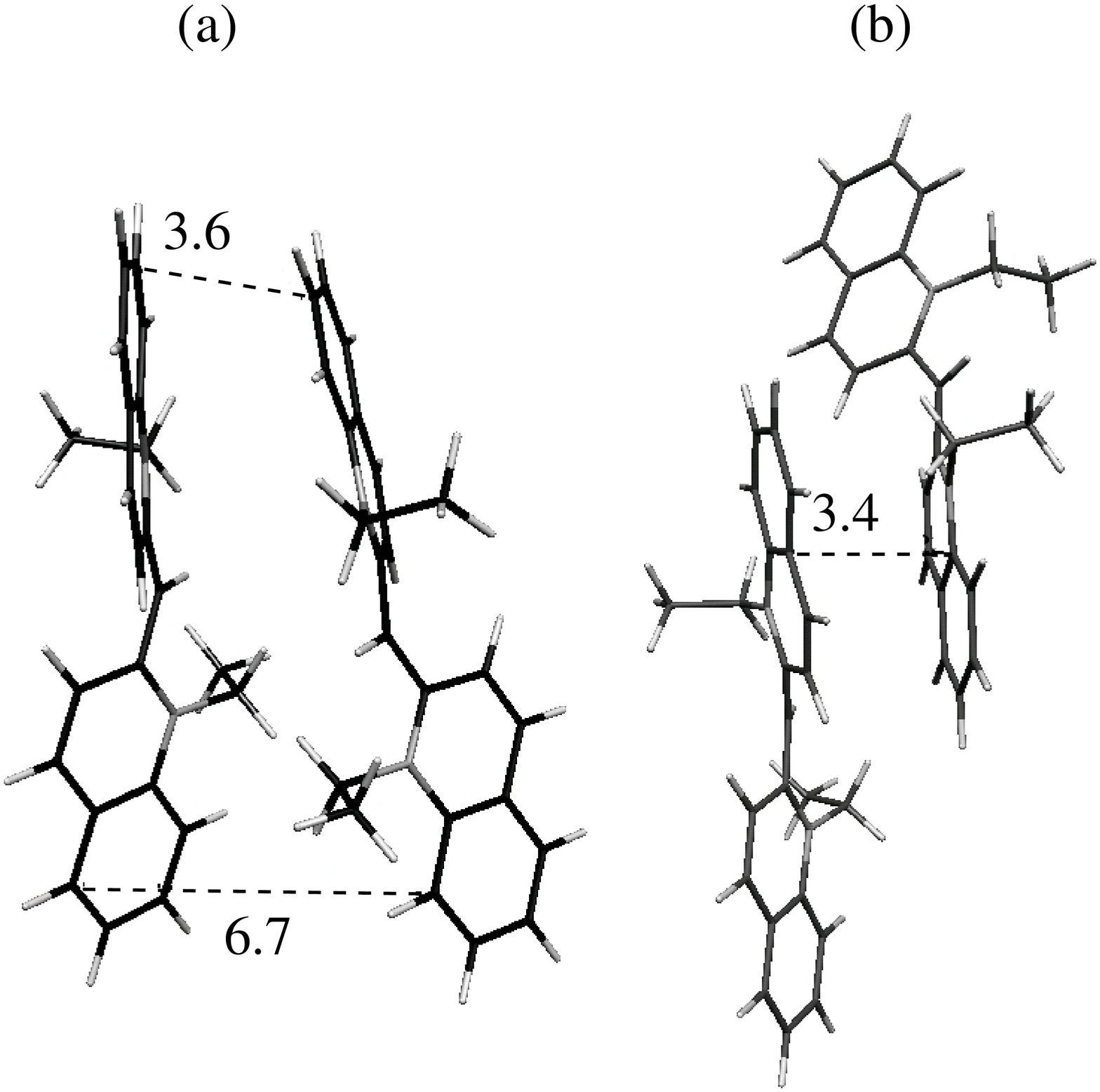}\end{center}
\end{figure}

\end{document}